\documentclass[twocolumn,showpacs,preprintnumbers,prb]{revtex4}
\usepackage{amssymb}
\usepackage{graphicx}
\usepackage{dcolumn}
\usepackage{bm}

\input{tcilatex}

\begin{document}

\title{Phase separation effects in charge-ordered Pr$_{0.5}$Ca$_{0.5}$MnO$%
_{3}$ thin film\\
}
\author{S. de Brion}
\author{G. Storch, G. Chouteau, A.\ Janossy}
\author{W. Prellier, E. Rauwel Buzin}
\date{\today}

\begin{abstract}
Compressed $Pr_{0.5}Ca_{0.5}MnO_{3}$ films (250nm) deposited on $LaAlO_{3}$
have been studied by Electron Spin Resonance technique under high frequency
and high magnetic field. We show evidences for the presence of a
ferromagnetic phase (FM) embedded in the charge-order phase (CO), in form of
thin layers which size depends on the strength and orientation of the
magnetic field (parallel or perpendicular to the substrate plane). This FM
phase presents an easy plane magnetic anisotropy with an anisotropy constant
100 times bigger than typical bulk values.\ When the magnetic field is
applied perpendicular to the substrate plane, the FM phase is strongly
coupled to the CO phase whereas for the parallel orientation it keeps an
independent ferromagnetic resonance even when the CO phase becomes
antiferromagnetic.
\end{abstract}

\maketitle
\preprint{EPJB}
\altaffiliation{Budapest University of Technology and Economics, Institut of physics, H1521
POB 91, Hungary}
\affiliation{Grenoble High Magnetic Field Laboratory, CNRS and MPI-FKF, B.P. 166, 38042
Grenoble Cedex 9, France}
\affiliation{CRISMAT, CNRS UMR 6508, ISMRA and Universit\'{e} de Caen, 6 boulevard du Mar%
\'{e}chal Juin, 14050 Caen Cedex, France}

% Force line breaks with \\




\section{\label{sec:level1}Introduction}

In the past few years the charge-ordering (CO)\ phenomena \cite{CO}
attracted much attention. This feature is observed in compounds like e.g. $%
Pr_{0.5}Ca_{0.5}MnO_{3}$ \cite{PCMObulk1,PCMObulk2} electrons become
localized due to the ordering of heterovalent cations in two different
sublattices with $Mn^{3+}$ and $Mn^{4+}$ions respectively.\ The materials
are insulating below the CO\ transition temperature (T$_{CO}$), but it is
possible to melt the CO state and render the material ferromagnetic and
metallic (FM state) by, for example, the application of a magnetic field. In 
$Pr_{0.5}Ca_{0.5}MnO_{3}$ this critical magnetic field \ is around 25T at
4K. One of the open questions which has attracted much attention recently is
the possibility of a coexistence of two phases \cite{PhaseSeparation} i.e.
the ferromagnetic metallic phase (FM) and the charge-order insulating (CO)
phase.

Few experiments are able to discriminate between a homogeneous magnetic
state and the averaged magnetic properties of a phase separated state. One
of them is Electron Spin Resonance (ESR)\cite{ESR}. We used this technique
recently to probe powder samples of the charge-order compound $%
Nd_{0.5}Ca_{0.5}MnO_{3}$, where no trace of the FM phase was detected \cite%
{NdCaESR}. In this case, a magnetic field higher than 15 Tesla is required
to destroy the CO state in favor of the FM state. The energy difference
between the CO phase and the FM phase is too large for the coexistence of
both. However, when the compound is in the form of a thin film deposited
onto SrTiO$_{3}$ or LaAlO$_{3}$ substrates, the substrate-induced strain
modifies the stability of both phases. In the absence of a magnetic field
the slightly incommensurate modulated structure that characterizes the
arrangement of $Mn^{3+}$ and $Mn^{4+}$ ions in the CO phase is modified: a
smaller incommensurate modulation vector than in the bulk has been observed
in $Pr_{0.5}Ca_{0.5}MnO_{3}$ thin films\ \cite{PrCAFilmGrowth}. Furthermore
the critical magnetic field required for the destruction of the CO phase is
also strongly reduced as compared to the bulk material: this was illustrated
recently in $Pr_{0.5}Ca_{0.5}MnO_{3}$ thin films \cite%
{PrCaFilm,PrCafilmthickness}. These properties of strained thin films make
them good candidates for the occurrence of phase separation and motivated
the present work. We have undertaken an Electron Spin Resonance (ESR) study
on 250 nm thick Pr$_{0.5}$Ca$_{0.5}$MnO$_{3}$ thin films (PCMO) grown on
LaAlO$_{3}$ (LAO). We present strong evidence for a phase separation, with
properties depending on the strength and the orientation of the applied
magnetic field (H) with respect to the substrate plane (parallel or
perpendicular).

\section{Experiments}

PCMO thin films were grown on (100)-LAO substrates using the pulsed laser
deposition technique.\ The details of the experimental procedure are
described elsewhere \cite{PrCAFilmGrowth}. We have shown that they are
single phase, $[101]$-oriented i.e. with the $[101]$ axis in the \textit{Pnma%
} space group being perpendicular to the substrate plane. In the pseudocubic
lattice, that will be further used for simplicity, this direction
corresponds to one of the quadratic axis i.e.$\ [001]$ .\ The out-of-plane
lattice parameter, at room temperature, for the 250 nm thick films is 0.384
nm confirming that the PCMO films are under compression in the plane of LAO
as previously shown \cite{PrCafilmthickness}. The stability of the CO\ phase
in the presence of magnetic field was studied by transport measurements \cite%
{PrCaFilm}.\ The transition towards the FM\ phase is shown by the dramatic
decrease of the electrical resistivity, the CO\ phase being insulating while
the FM state is metallic.\ The corresponding phase diagram is given in
Figure 1 together with typical magnetoresistance curves. The CO/FM
transition as a function of magnetic field is dramatically reduced compared
to the bulk when the temperature is lowered. At 150K e.g. it is about 8
Tesla\ in\ the film and 15 Tesla in the bulk \cite{PCMObulk2}. The
difference in thermal dilatations between the film and the substrate
destabilizes the CO phase. Note also the first order character of the \
transition evidenced by the hysteresis as previously observed in CO
compounds with other compositions \cite{NSMO}.

The ESR spectra were recorded at 9.44 GHz using a conventional Brucker
spectrometer operating with a 1 Tesla electromagnet.\ The thin film was cut
into a 2x4 mm$^{2}$ plate to avoid cavity saturation, in the paramagnetic
regime at least.\ For the high field, high frequency measurements, the whole
film was used (two 2x4mm$^{2}$ plates superimposed).\ For these experiments
a homemade spectrometer \cite{ESRspectro} was used \ at frequencies of 95
GHz and 190 GHz with a superconducting magnet operating up to 12 Tesla. The
magnetic field was kept below the CO/FM\ transition (6 Tesla for 95 GHz and
8 Tesla for 190 GHz recorded at 150 K) in order to test the phase separation
scenario without passing the first order, hysteretic transition. The high
frequency spectrometer uses oversized cylindrical guides for the
electromagnetic wave with no cavity at all. This facilitates measurements in
the very large frequency range, but the sensitivity is rather poor. Up to
four spectra had to be \ recorded and averaged at each temperature.

\section{\protect\bigskip X band measurements}

Figure 2 reports on X band spectra for H parallel to the film plane. The
spectra were obtained starting from low temperatures. No noticeable change
appears with thermal cycling.\ We checked that the sharp features correspond
to the LAO\ substrate and the signal close to 0.9 Tesla is due to air in the
cavity and it disappears under nitrogen pressure. The broad signal at $g=2.0$
arises from the film itself.\ It depends strongly on temperature but very
weakly on field orientation as can be seen from Figure 3 where the line
position $B_{0}$, intensity (proportional to the ESR susceptibility $\chi $)
and peak to peak line width $\Delta B_{0}$ are plotted. Below 150-160K, the
line intensity is weakened and deviates markedly from a Lorenzian shape and
it finally vanishes.\ We attribute these effects to the opening of the
antiferromagnetic gap and deduce the N\'{e}el temperature: $T_{N}\simeq 140K$%
, a value lower than the reported value for the bulk $(180K)$ \cite%
{LaCaMnOphase diagram}. The ESR susceptibility follows the general trend of
the magnetic susceptibility with a maximum at $T_{CO}$ identified here at $%
T_{CO}\simeq 240-250K$ , which corresponds also to the minimum in the ESR
linewidth. This value is very close to the one deduced from resistivity
measurements on a similar film as well as to the bulk value indicating that $%
T_{CO}$ does not seem to depend on the\ film thickness and the
substrate-induced stress.\ However when the temperature is lowered, the
changes in cell parameters observed in bulk materials is quenched due to the
substrate stress and the CO state has an incommensurate modulated structure 
\cite{PrCAFilmGrowth}.\ This affects also the magnetic exchange interactions
between manganese ions and the antiferromagnetic ordering temperature
compared to the bulk is reduced.

The ESR line position B$_{0}$ is rather constant in the temperature range
200 K- 300 K. The slight variation as a function of field orientation (6 mT
at 200 K) arises from demagnetization effects. The resonant condition is
given by $H_{res}=\sqrt{\left( H_{0}+(n_{y}-n_{z})M\right) \left(
H_{0}+(n_{x}-n_{z}\right) M)}$ \cite{Herpin}\ where $H_{0}$ is the applied
magnetic field in the z direction, $M$ the sample magnetization, $n_{x,y,z}$
the demagnetization factors where the z axis is along the magnetic field
direction and $H_{res}$ is the resonance field for a given electromagnetic
wave frequency $\nu :$ $h\upsilon =g\mu _{B}H_{res}$, $g$ is the
gyromagnetic factor. Taking the infinite plate geometry and a magnetization $%
M=2.10^{-3}\mu _{B}$ per unit formula at 200 K, the correction for the field
applied parallel to the plate is $-2mT$ and for the perpendicular direction $%
+4mT$ . These corrections account well for the discrepancy in line positions
for the parallel and perpendicular field directions. We thus conclude that
the g factor is equal to 2.02(1) with no crystalline anisotropy.

\section{High frequency ESR measurements}

The high frequency, high field ESR\ measurements enable us to get more
detailed information on the local magnetic order. Spectra recorded at 95 GHz
and 190 GHz are presented in figure 4 and 5 for the static field applied
parallel to the film plane and in figure 6 for the perpendicular direction.\
The sharp features at $g=2.0$ are due to the substrate.\ Note the decrease
in sensitivity as compared to the X band measurements.\ Nevertheless we are
able to detect the manganite film itself which is clearly identified as the
broad signal similarly to the X band results. For both field directions, the
paramagnetic signal of the CO phase disappears below 130 K due to the
opening of the antiferromagnetic gap as already shown in the X band
measurements.\ More surprisingly, another signal is present, which depends
on the temperature and the applied magnetic field.\ For the field parallel
to the film plane, this additional signal is on the low field side of the
paramagnetic line with the same shift at 95 GHz and 190 GHz (around 0.8T):
it is characteristic of a ferromagnetic resonance. It is well defined below
200 K and persists below $T_{N}$, revealing the presence of a segregated
ferromagnetic phase, weakly coupled to the CO phase. For the field applied
perpendicular to the film plane, this additional signal is on the high field
side of the paramagnetic line; it is not as well defined and it spreads
continuously from the paramagnetic line up to a maximum shift is of about
1.6T. It disappears below $T_{N}$ , this is expected if the segregated
ferromagnetic phase is coupled to the CO phase. In this case the\
ferromagnetic regions couple to the AF regions and the ferromagnetic
resonance mode shifts and broadens if the ferro regions are small.\ 

To model the ferromagnetic resonance, we include in addition to the applied
field \ $H_{0}$, the demagnetization field and the anisotropy field which
act on the rotating part of the magnetization induced by the electromagnetic
wave. Assuming homogeneous fields within the ferromagnetic regions, we use
the concept of demagnetization factors and the resonant condition becomes: $%
H_{res}=\sqrt{\left( H_{0}+(n_{y}+n_{ay}-n_{z})M\right) \left(
H_{0}+(n_{x}+n_{ax}-n_{z}\right) M)}$ \cite{Herpin} where $n_{ax,ay}$ denote
the fictitious anisotropy constants. For instance, in a cubic structure with 
$H_{0}$ parallel to one of the cubic axis $n_{ax}=n_{ay}=2K/M^{2}$ where $M$
is the ferromagnetic magnetization and $K$\ the anisotropy constant.\ If $K$
is positive, the cubic axis is the easy magnetic axis; if $K$ is negative,
it is the hard axis.\ At high fields, the resonant condition reduces to $%
H_{res}=H_{0}+(n_{x}+n_{y}+n_{ax}+n_{ay}-2n_{z})M/2$.\ 

We first consider demagnetization effects. We may assume that the shape of
the segregated phase is isotropic in the film plane. Then only one
demagnetization factor is needed, $n_{//}$ for instance, which relates to a
direction in the film plane. When the field is applied perpendicular to the
film, the resonance condition is $H_{res}=H_{0}-(1-3n_{//})M$. For a
measurement at fixed frequency, this leads to a shift $\mu _{0}\Delta H=$ $%
\mu _{0}(1-3n_{//})M$. This shift will be positive and maximum for a plate
like shape \ ( $n_{//}=0$ and $\mu _{0}\Delta H=$ $\mu _{0}M$); it will be
negative for a needle like shape\ ($n_{//}=1/2$ and $\mu _{0}\Delta H=$ $%
-\mu _{0}M/2$). The observed shift at 140 K, $\mu _{0}\Delta H_{obs}=+1.53T$%
, is positive and even larger than $\mu _{0}M$ (the saturated value at low
temperatures is about 0.8T). This is in favor of a plate like shape with an
additional shift arising from anisotropy effects. Since $H_{0}//\left[ 001%
\right] $ of the pseudo cubic cell, $n_{a}=2K/M^{2}$and $\mu _{0}\Delta
H=\mu _{0}(M$ $-2K/M)$. At 140 K, assuming \ $\mu _{0}M=0.7T$ close to the
saturation ferromagnetic value, then $\mu _{0}H_{a}=\mu _{0}2K/M=-0.8T$, the
same order of magnitude as $\mu _{0}M$. Such a strong anisotropy field is
not usual in manganese perovskites: \ the corresponding anisotropy constant (%
$K\simeq -2.10^{5}J/m^{3}$) \ is 100 times larger than in bulk La(Sr,Pb)MnO$%
_{3}$ crystals \cite{Perekalina} and is of opposite sign.\ In our thin film
the FM phase embedded in the CO phase has an enhanced magnetic anisotropy as
compared to bulk ferromagnetic samples and contrary to bulk samples the $%
\left[ 001\right] $ direction is a hard magnetic axis.

For the field applied in the plane of the substrate, $%
H_{res}=H_{0}+(1-3n_{//})M/2$ if demagnetization effects alone are taken
into account, leading to a shift $\mu _{0}\Delta H=$ $\mu _{0}(3n_{//}-1)M/2$%
.\ The observed shift at 140K$\ $is $\mu _{0}\Delta H_{obs}=-0.86T$. The
sign of the shift for this field orientation also suggests that the
ferromagnetic phase consists of plate like regions with $n_{//}\simeq 0$.\
Anisotropy effects have to be included here also and these\ give rise to an
additional negative shift: $\mu _{0}\Delta H=-\mu _{0}(M$ $+2K/M)/2$ where $%
K $\ appears to be positive this time.\ Since the film is twinned in the $%
\left[ 010\right] $ and $\left[ 010\right] $ directions, a positive
anisotropy constant means that both these axis are easy magnetic axis, i.e.
the FM phase has an easy plane anisotropy.\ For an infinitely thin plate
like shape,\ at high fields in the plane, the shift is half that appearing
for a field applied perpendicular to the film plane. Figure 7\ presents, in
a frequency-field diagram, the experimental resonant fields together with
the calculated contribution arising from demagnetization effects (assuming a
plate like shape for the ferromagnetic phase) and the one arising from
anisotropy effects (assuming easy plane anisotropy). A sketch of the
segregated ferromagnetic phase is also drawn with the direction of the
magnetic easy axis.

Our results should be compared to observations in very thin ferromagnetic
films where the strain from the substrate is important \cite{Bosovic}, \cite%
{easyAxisFilm}. In such cases the positive magnetostriction increases the
anisotropy constants: the compressive strain in films grown on LAO\
substrates results in an out of plane easy magnetic axis while growth on STO
substrates, which induces tensile strain, induces an in-plane easy magnetic
axis. The enhanced anisotropy of the ferromagnetic phase embedded in the CO
phase in our PCMO film can then be understood by the stress induced by the
CO phase. Similarly to what happens in thin films, the magnetic anisotropy
will be enhanced with larger crystallographic mismatch between the two
phases and thinner ferromagnetic layers. The experiments suggest that the FM
phase grows in form of thin layers, as was already suggested by the
demagnetization effects, and that the FM\ phase is compressed by the CO
phase.

Figure 8 shows the temperature variation of $M+2\left\vert K\right\vert /M$
calculated from all the measurements: at 95 GHz ad 190 GHz for $H_{0}$
parallel to the substrate plane, and at 95 GHz for the perpendicular
direction. The different values agree quite well, confirming our model of a
plate like shape with easy plane anisotropy. We note a slight increase of $%
M+2\left\vert K\right\vert /M$ with the magnetic field at which the
resonance occurs for a given temperature.\ At higher field, the segregated
phase grows, the strain decreases and $K$\ becomes smaller. At the same
time, the magnetization $M$ increases so that the two effects add. As for
the dependence on the field orientation, the segregated phase is much better
defined when the field is applied perpendicular to the film plane. This can
be seen from the shape of the ESR spectrum arising from the FM\ phase: it is
a well resolved line for the perpendicular direction while it extends
continuously from the paramagnetic CO signal for the parallel direction. The
latter spectrum indicates a distribution in the shape factor and anisotropy
constant. Moreover, the FM line persists at temperatures below the
antiferromagnetic transition of the CO phase for the field perpendicular to
the film plane, but it disappears below the antiferromagnetic transition for
the parallel direction.

Note also that no trace of this segregated phase was detected in the X band
measurements on a similar film: this was checked at 200 K and 100 K where we
looked at the full angular dependence of the ESR\ spectrum in the field
range 0-1T.\ This confirms that the amount of this segregated phase depends
on the strength of the magnetic field. At fields up to 7T, at 140 K for
instance, this FM\ phase is still a minority phase.

In conclusion, we have shown that a segregated ferromagnetic phase grows
within the CO phase in a 250 nm thick PCMO/LAO\ film. The segregated phase
has the shape of very thin layers parallel to the film plane. The segregated
domains grow with increasing external field.\ The segregated FM phase has an
enhanced anisotropy constant with an easy plane type of anisotropy.\ It also
depends on the orientation of the applied magnetic field: the segregated
phase is better defined for H parallel to the film. In the perpendicular
direction, the ferromagnetic resonance disappears below the
antiferromagnetic ordering temperature of the CO\ phase due to a strong
magnetic coupling.\ This effect may explain why no segregated phase was
observed by ESR\ in LaSrMnO crystals \cite{Pimenov}, while neutron
measurements were able to detect it \cite{Moussa}.

We are indebted to J. Dumas for the X band ESR\ measurements. The Grenoble
High Magnetic Field Laboratory is 'laboratoire associ\'{e} \`{a} l'Universit%
\'{e} Joseph Fourier-Grenoble'. AJ acknowledges the support of the Hungarian
state grant OTKA T043255.

%% \section{Figure captions}

%Just because of unusual number of tables stacked at end
\bibliographystyle{plain}
\bibliography{apssamp}
% Produces the bibliography via BibTeX.

\clearpage
\begin{figure}
    \centering
   \vspace{1cm}
\includegraphics[width=14cm]{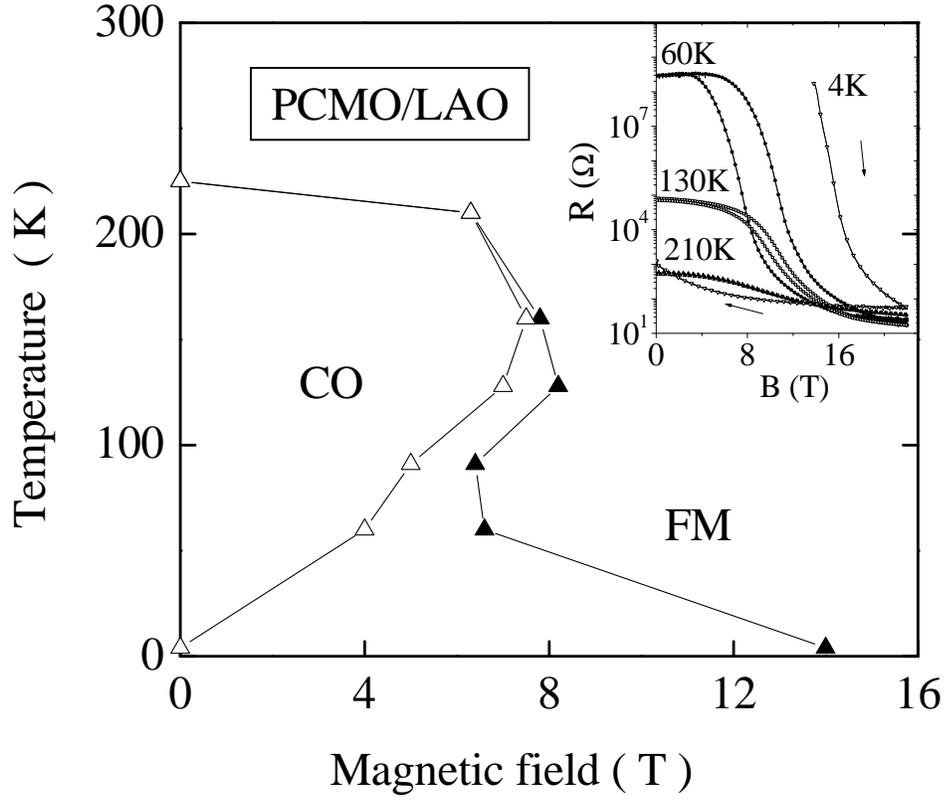}
\caption {Temperature, magnetic field phase diagram of a 250 nm thick Pr$_{0.5}$Ca$_{0.5}MnO_{3}$ thin film deposited on LaAlO$_{3}$ for
increasing ($\blacktriangle $) and decreasing ($\Delta $) field. Inset:
magnetoresistance curves at different temperatures.}
 \label{Figure 1}
 \end{figure}

\clearpage

\begin{figure}
    \centering
  \vspace{1cm}
\includegraphics[width=14cm]{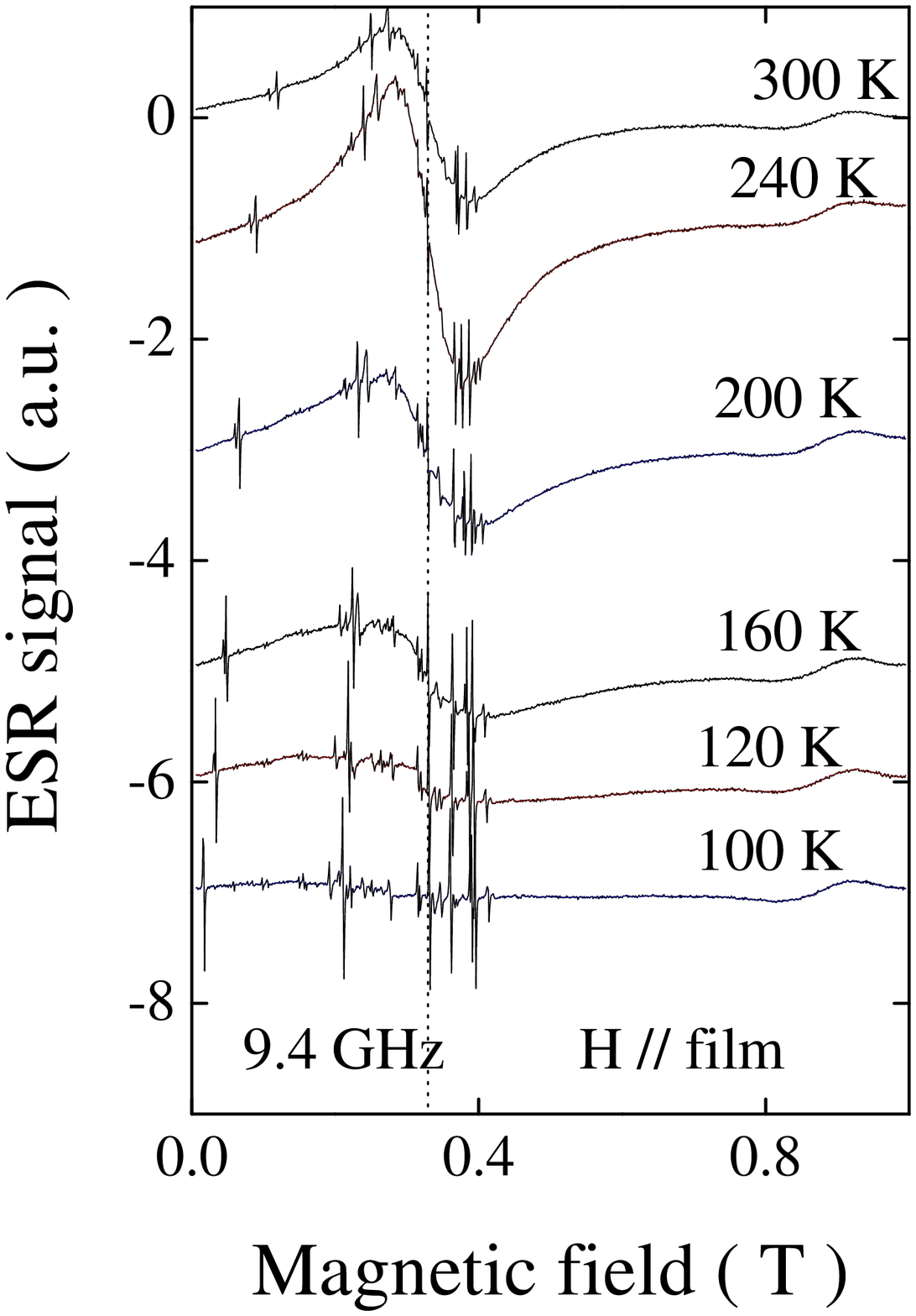}
\caption{ESR spectra taken at 9.44 GHz in PCMO/LAO, the static magnetic field was applied parallel to the film plane. The dotted line
corresponds to g=2.00. The sharp features arise from the LAO substrate and
the large signal around g=2.00 from the PCMO film.}
 \label{Figure 2}
 \end{figure}

\clearpage

\begin{figure}
    \centering
  \vspace{1cm}
\includegraphics[width=14cm]{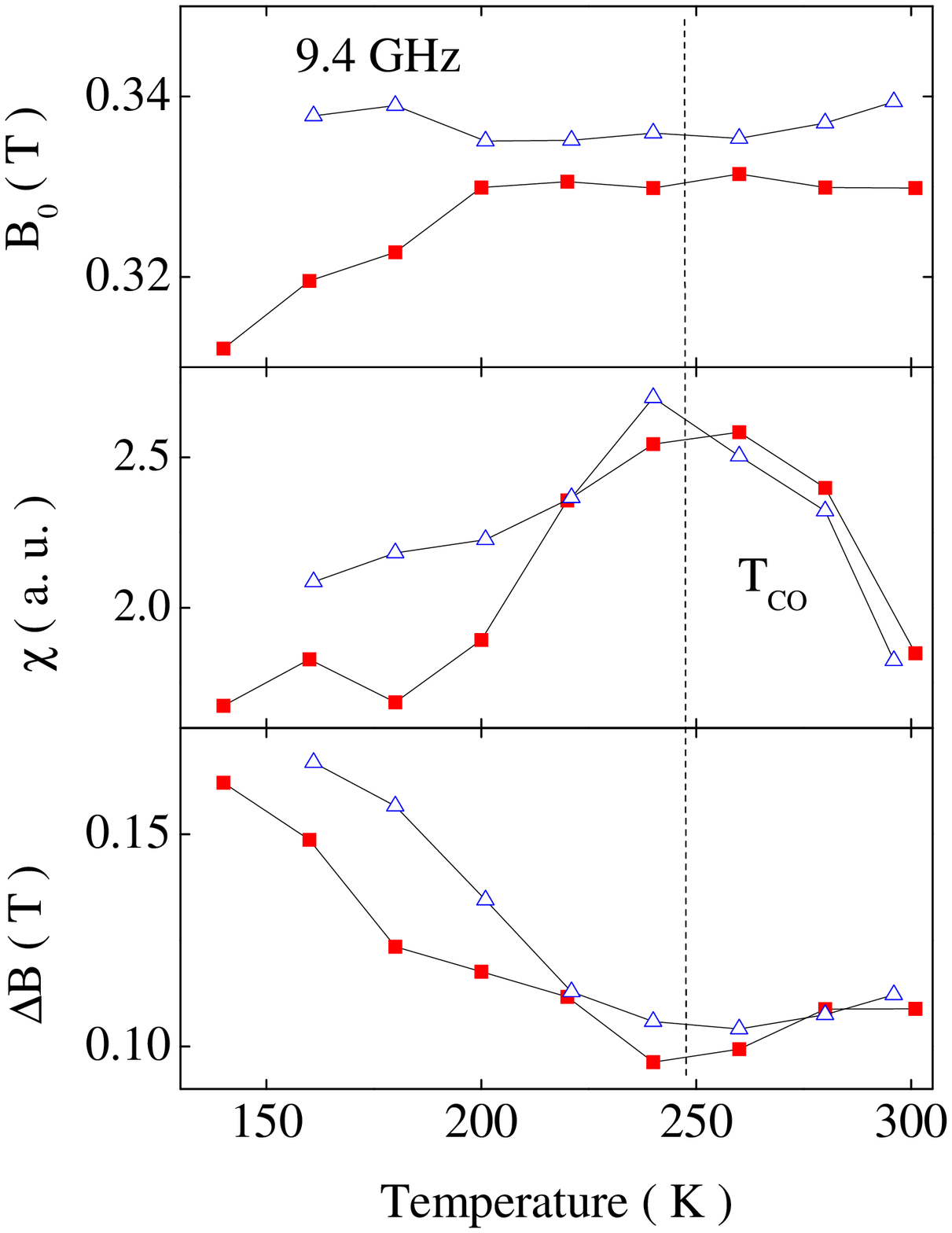}
\caption{ESR line position, intensity and line width as a
function of temperature in $PCMO/LAO$ taken at $9.44GHz$ for the magnetic
field applied parallel ($\blacksquare $) and perpendicular ($\Delta $) to
the film plane.}
 \label{Figure 3}
 \end{figure}

\clearpage

\begin{figure}
    \centering
  \vspace{1cm}
\includegraphics[width=14cm]{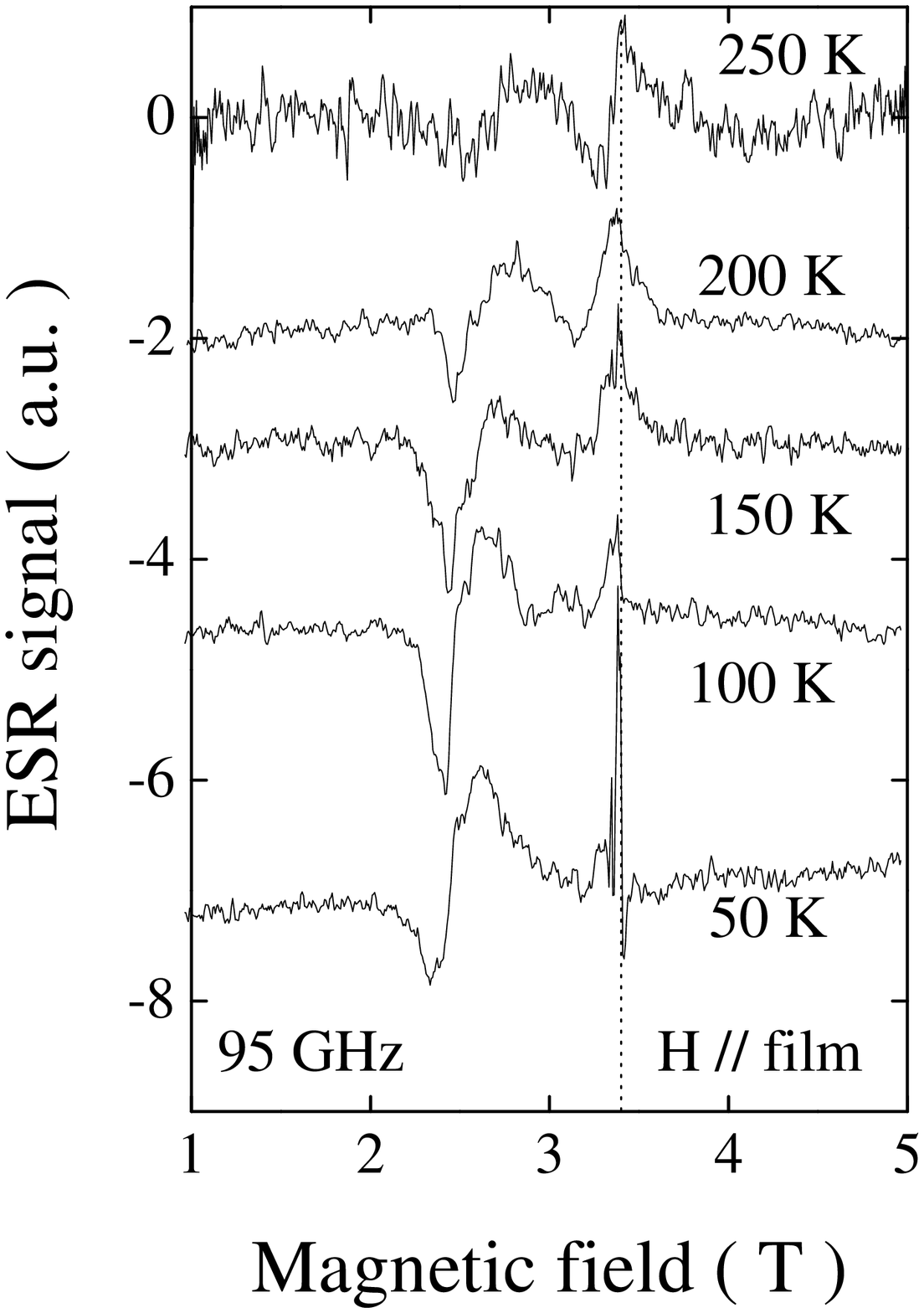}
\caption{ESR spectra as a function of magnetic field in $PCMO/LAO$ taken at $95GHz$ and at different temperatures. The magnetic
field was applied parallel to the film plane. The dotted line corresponds to
g=2.00.}
 \label{Figure 4}
 \end{figure}

\clearpage

\begin{figure}
    \centering
  \vspace{1cm}
\includegraphics[width=14cm]{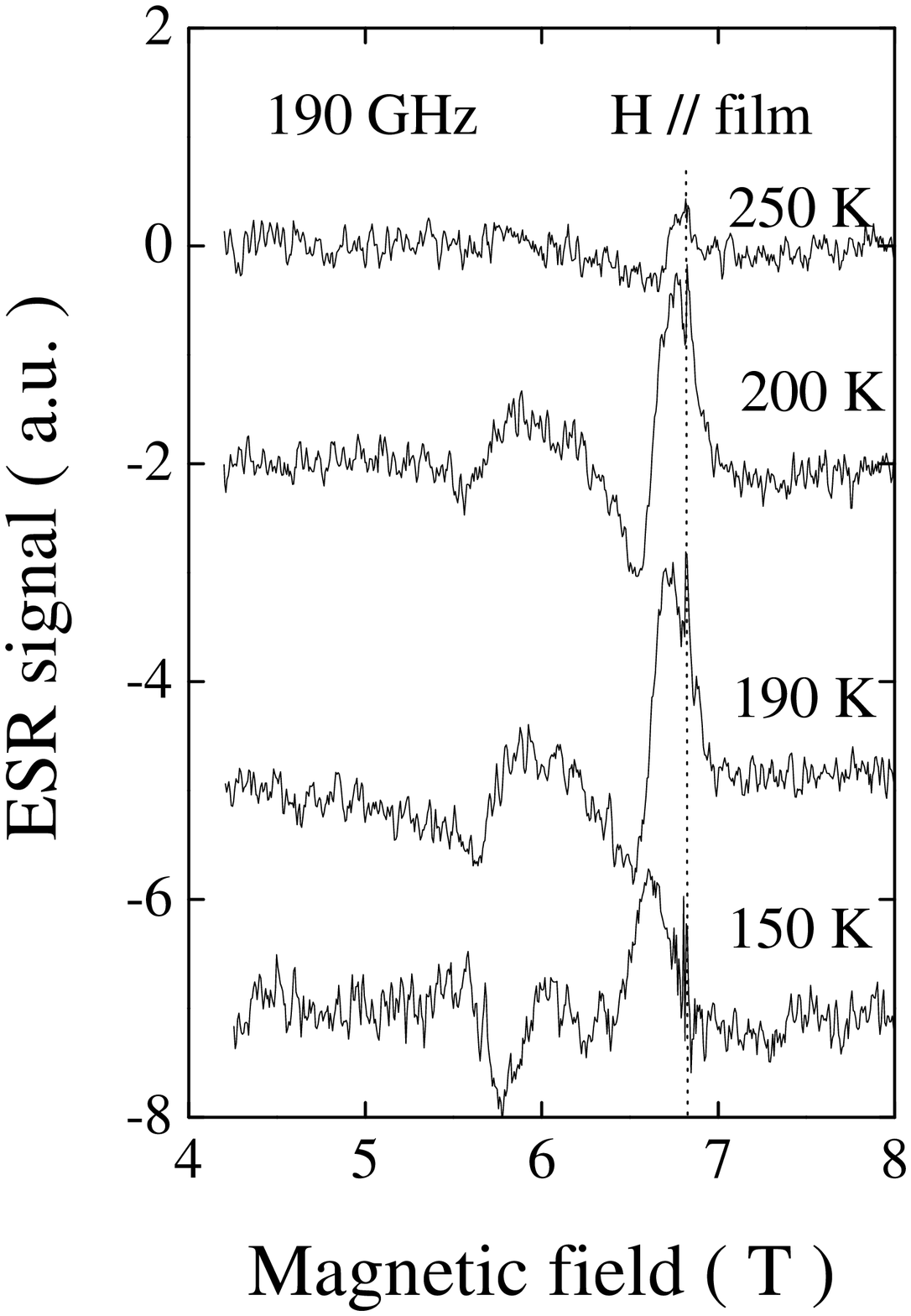}
 \caption{ESR spectra as a function of magnetic field in $PCMO/LAO$ taken at $190GHz$ and at different temperatures. The magnetic field was applied parallel to the film plane. The dotted line corresponds to
g=2.00.}
\label{Figure 5}
 \end{figure}

\clearpage

\begin{figure}
    \centering
  \vspace{1cm}
\includegraphics[width=14cm]{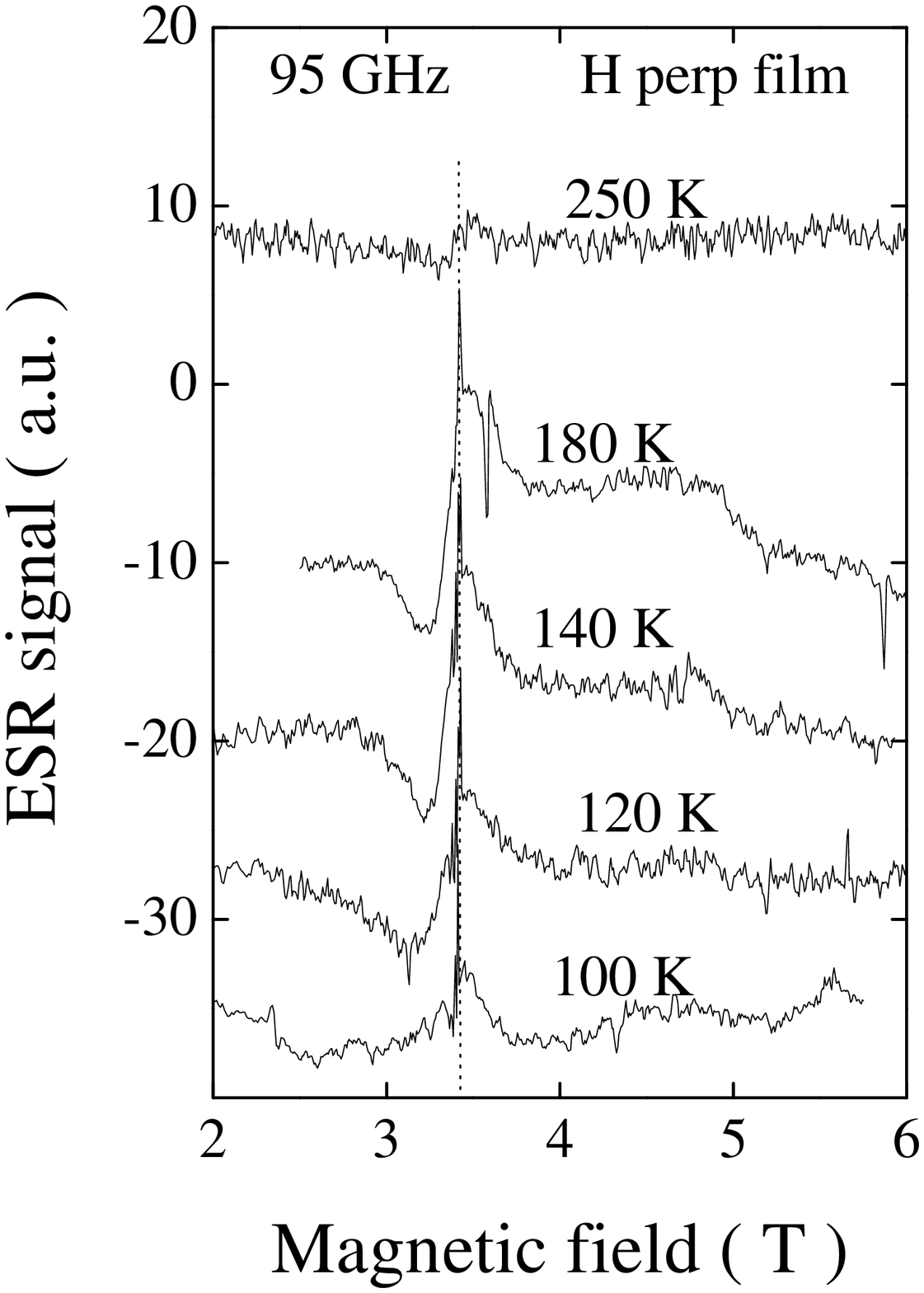}
\caption{ESR spectra as a function of magnetic field in 
$PCMO/LAO$ taken at $95GHz$ and different temperatures. The magnetic field was applied perpendicular to the film plane. The dotted line corresponds to g=2.00.}
 \label{Figure 6}
 \end{figure}

\clearpage

\begin{figure}
    \centering
  \vspace{1cm}
\includegraphics[angle=-90, width=14cm]{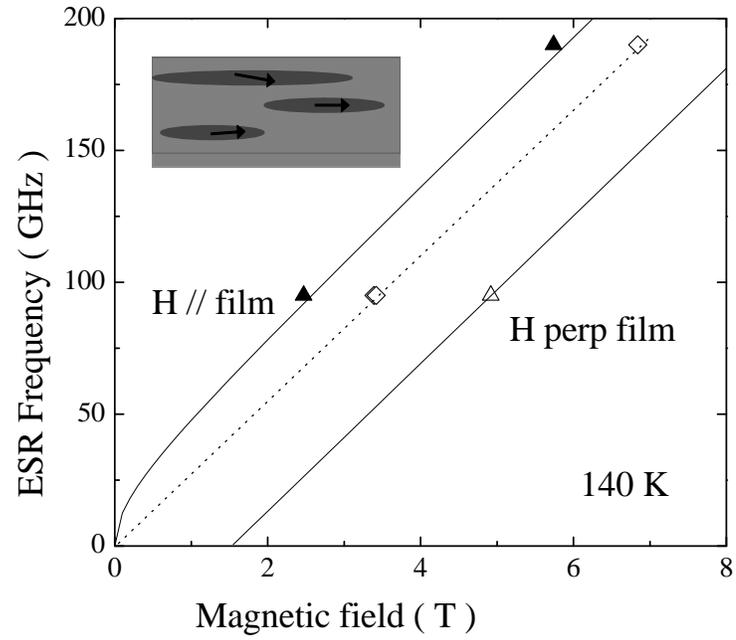}
\caption{Ferromagnetic resonance modes in a frequency-field
diagram: experimental values at 140K for the magnetic field applied parallel
($\blacktriangle $) and perpendicular ($\Delta $) to the film plane and
calculated values using the model described in the text (continuous lines).
The paramagnetic resonance in the CO phase ($\diamond )$ and calculated
(doted line) are also shown.}
 \label{Figure 7}
 \end{figure}

\clearpage

\begin{figure}
    \centering
  \vspace{1cm}
\includegraphics[width=14cm]{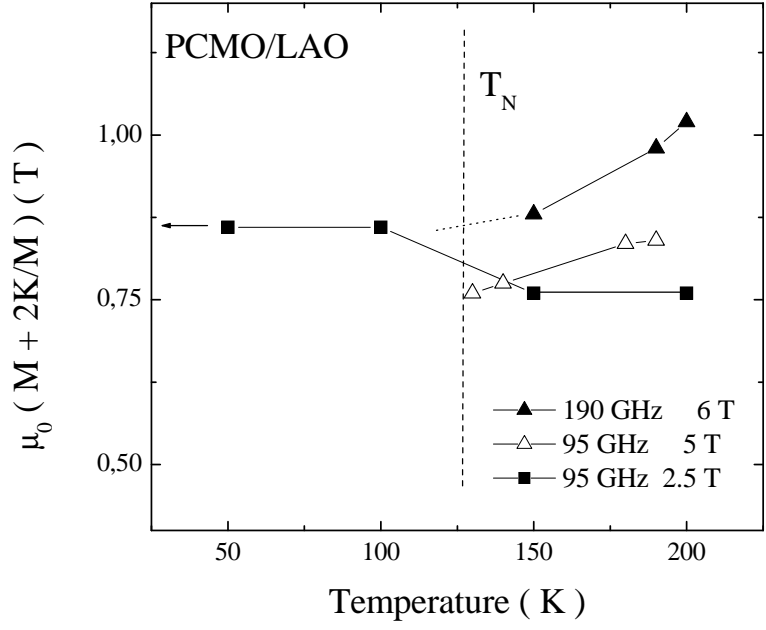}
\caption{Temperature dependence of $M+2\left\vert K\right\vert /M^{2}$ calculated from the ESR line shifts (see main text) as a function
of temperature for the magnetic field applied parallel ($\blacksquare,\blacktriangle $) and perpendicular ($\Delta $) to the film plane.}
 \label{Figure 8}
 \end{figure}

\end{document}